\documentclass[aps,preprint]{revtex4}%
\usepackage{amsfonts}
\usepackage{amsmath}
\usepackage{amssymb}
\usepackage{graphicx}
\usepackage{natbib}
\usepackage{chapterbib}
\providecommand{\U}[1]{\protect\rule{.1in}{.1in}}

\begin{document}
\bibliographystyle{unsrtnat}
\preprint{ }
\title{Classical Representation of a Quantum System at Equilibrium: Applications }
\author{Sandipan Dutta and James Dufty }
\affiliation{Department of Physics, University of Florida}

\begin{abstract}
In the preceding paper, the structure and thermodynamics of a given
quantum system was represented by a corresponding classical system having
an effective temperature, local chemical potential, and pair
potential. Here, that formal correspondence is implemented
approximately for applications to two quantum systems. The first is
the electron gas (jellium) over a range of temperatures and
densities. The second is an investigation of quantum effects on
shell structure for charges confined by a harmonic potential.

\end{abstract}
\date{01 Feb 2013}
\maketitle

\section{Introduction}

\label{sec1}In a companion paper \cite{DD12} a method was described
that would allow application of strong coupling classical many body
methods to calculate properties of equilibrium quantum systems.
Within the grand ensemble for equilibrium statistical mechanics, the
thermodynamics and structure are obtained as functions of the
temperature $T$, the local chemical potential $\mu\left(
\mathbf{r}\right)  \equiv\mu-\phi_{ext}\left( \mathbf{r}\right) $,
and a pair potential $\phi(\mathbf{r},\mathbf{r}^{\prime})$, (where
$\phi_{ext}\left( \mathbf{r}\right)  $ is an external single
particle potential). A grand ensemble for a corresponding classical
system is characterized by an effective temperature $T_{c}$, an
effective local
chemical potential $\mu_{c}\left(  \mathbf{r}\right)  \equiv\mu_{c}%
-\phi_{c,ext}\left(  \mathbf{r}\right)  $, and an effective pair
potential $\phi_{c}(\mathbf{r},\mathbf{r}^{\prime})$. These three
classical parameters are fixed by three correspondence conditions:
equivalence of classical and quantum pressures, densities, and pair
correlation functions. An approximate inversion of these formal
definitions to obtain $T_{c}$, $\mu_{c}\left(
\mathbf{r}\right)  $, and $\phi_{c}(\mathbf{r},\mathbf{r}%
^{\prime})$ was described within classical liquid state theory. The
objective here is to illustrate this approach for two applications.
The first is to calculate the pair correlation functions for the
electron gas (jellium), the prototypical test bed for quantum
correlations \cite{KSK05,Vignale} , over a wide range of
temperatures and densities. Corresponding thermodynamic properties
can then be calculated in terms of these correlation functions. The
second application is to harmonically bound charges in a trap
\cite{Wrighton09}. Specifically, the role of quantum diffraction and
exchange as a mechanism for shell formation is investigated. While
both systems have been studied extensively at both very low (ground
state) and very high (plasma) temperatures, the relevance here is a
method that applies across the intermediate domain.

The use of effective pair potentials to include some quantum effects
in classical methods like molecular dynamics simulation has a long
history \cite{Jones07}. A new phenomenological approach proposed by
Perrot and Dharma-wardana \cite{DWP00} more recently goes a step
farther to introduce an effective classical temperature as well.
Applications of this extended approach to a variety of systems and
properties over the past decade have met with remarkable success
\cite{DW11}. The present work can be considered as a parameter free
formalization of this earlier work, and comparisons are discussed
critically here as well.

In the next section, the approach of reference \cite{DD12} is
applied to jellium. The effective classical pair potential is
discussed and illustrated, and then the pair correlation function is
calculated using the classical strong coupling hypernetted chain
integral equations (HNC) \cite {Hansen}. The dimensionless
temperature relative to the Fermi temperature, $t=T/T_{F}$,  is
considered in the range
$0\leq t\leq10$. The dimensionless measure of the density is $r_{s}%
=r_{0}/a_{B}$, the mean distance between particles defined by $4\pi r_{0}%
^{3}/3=1/n$ relative to the Bohr radius $a_{B}$. It is considered in
the range $1\leq r_{s}\leq10$. The results are compared to the above
mentioned model of Perrot and Dhama-wardana (PDW), and to the
extension  of the self-consistent generalization of the random phase
approximation, STLS \cite{STLS70}, to finite temperatures by Tanaka
and Ichimaru \cite{Tanaka86}. Comparisons with limited diffusion
Monte Carlo simulation data for $t=0$ \cite{Ortiz} and recent
restricted path integral Monte Carlo simulations results for finite
temperatures \cite{Brown} show good agreement.

A second application is described briefly in Section \ref{sec6}, to
charges in a harmonic trap where classical strong Coulomb
correlations produce shell structure \cite{Wrighton09}.  These
results are of interest for laser cooled ions in traps, where the
quantum effects are expected to be weak, and for electrons in
quantum dots at low temperatures, where the quantum effects are
strong. In the classical case, strong Coulomb correlations are
required for shell structure - they are absent in a mean field
theory at any value of the coupling constant. The objective here is
to explore possible new origins for shell structure due to quantum
effects. It is shown that the quantum mean field theory, without any
Coulomb correlations, leads to shell structure due to diffraction
effects modifying the Coulomb interactions and/or exchange effects
modifying the effective trap potential.

The results are summarized and discussed in the last section.

\section{Application to Uniform Electron Gas}

\label{sec2}The interacting electron gas is a one component system
of charges with Coulomb interactions embedded in a uniform
neutralizing back ground. The uniform electron gas, or "jellium",
provides an important model system to discuss correlations and
quantum effects in real physical metals, solids, and plasmas
\cite{KSK05}. The classical limit is known as the one component
plasma. There are two parts to this section. First, the parameters
$\beta
_{c}=1/k_{B}T_{c}$, $\mu_{c}(\mathbf{r}),$ and $\phi_{c}(\mathbf{r}%
,\mathbf{r}^{\prime})$ for the effective classical system are
determined approximately from their definitions in reference
\cite{DD12} (for the uniform
jellium $\mu_{c}(\mathbf{r})\rightarrow\mu_{c}$ and $\phi_{c}(\mathbf{r}%
,\mathbf{r}^{\prime})\rightarrow\phi_{c}(\left\vert \mathbf{r}-\mathbf{r}%
^{\prime}\right\vert )$). Second the resulting classical system is
applied to calculate the structure and thermodynamics of jellium
from the classical HNC liquid state theory.

To summarize the results of \cite{DD12}, the classical parameters
are defined as following. The classical temperature is obtained from
the correspondence condition of equal pressures and the classical
virial equation
\begin{equation}
\frac{\beta_{c}}{\beta}=\frac{\beta_{c}p_{c}}{\beta
p}=\frac{n}{\beta
p}\left[  1-\frac{n}{6}\int d\mathbf{r}h(r)\mathbf{r}\cdot\nabla\beta_{c}%
\phi_{c}(r)\right]  .\label{5.1}%
\end{equation}
The replacement of the pair correlation function $g(r)$ by the hole
function $h(r)=g(r)-1$ occurs because of the uniform neutralizing
background. The classical activity $\beta_{c}\mu_{c}$ is given by \cite {Baus}%
\begin{equation}
\beta_{c}\mu_{c}=\ln\left(  n_{c}\lambda_{c}^{3}\right)  -n\int d\mathbf{r}%
\left(  c(r)+\beta_{c}\phi_{c}(r)-\frac{1}{2}h(r)\left(
h(r)-c(r)\right)
\right)  ,\label{5.2}%
\end{equation}
where $c(r)$ is the direct correlation function defined in terms of
$h(r)$ by the Ornstein-Zernicke equation \cite {Hansen}%
\begin{equation}
c\left(  r\right)  =h\left(  r\right)  -n\int
d\mathbf{r}^{\prime}c(\left\vert
\mathbf{r-r}^{\prime}\right\vert )h\left(  r^{\prime}\right)  .\label{5.4}%
\end{equation}
Finally, the pair potential is obtained from the inversion of the HNC equation%
\begin{equation}
\beta_{c}\phi_{c}(r)=-\ln\left(  1+h(r)\right)  +h\left(  r\right)
-c\left(
r\right) . \label{5.3}%
\end{equation}
The classical pair correlation functions on the right
sides of these expressions have been replaced by the quantum
functions, according to the third correspondence condition. Hence
these classical parameters are determined by quantum input.

The practical approach is to provide the essential quantum input by
specifying $h(r)$ in some approximation. Equations (\ref{5.3}) and
(\ref{5.4}) then determine $\beta_{c}\phi_{c}(r)$ and $c\left(
r\right)  $, and with these known $\beta_{c}/\beta$ and
$\beta_{c}\mu_{c}$ can be calculated. The objective here is to
propose a simple approximation for practical application.

\subsubsection{Classical potential $\beta_{c}\phi_{c}(r)$}

The dominant exchange effects are already present in the ideal gas
calculation described in reference \cite{DD12}. Therefore it is
convenient to write
$\beta_{c}\phi_{c}(r)$ in the form%
\begin{equation}
\beta_{c}\phi_{c}(r)=\left(  \phi_{c}(r)\right)  ^{(0)}+\Delta(r),\label{8.1}%
\end{equation}
where $\left(  \phi_{c}(r)\right)  ^{(0)}$ is the ideal gas Pauli
potential and $\Delta(r)$ denotes the contribution to the effective
potential from the Coulomb interactions. In the classical limit
$\Delta(r)\rightarrow\beta q^{2}/r$. Another exact limit is the weak
coupling limit for which the direct correlation function becomes
proportional to the potential, or stated inversely,
\begin{equation}
\beta_{c}\phi_{c}(r)\rightarrow-c(r),\hspace{0.2in}\left(
\beta_{c}\phi
_{c}(r)\right)  ^{(0)}\rightarrow-c^{(0)}(r).\label{8.2}%
\end{equation}
Thus a possible approximation incorporating this limit is%
\begin{equation}
\beta_{c}\phi_{c}(r)\rightarrow\left(  \beta_{c}\phi_{c}(r)\right)
^{(0)}-\left(  c(r)-c^{(0)}(r)\right)  ^{(w)},\label{8.3}%
\end{equation}
where $\left(  c(r)-c^{(0)}(r)\right)  ^{(w)}$ denotes a weak
coupling calculation of the direct correlation functions from the
Ornstein - Zernicke equation (\ref{5.4}). For the classical OCP
(Coulomb potential) this yields the Debye - Huckel approximation to
$h\left(  r\right)  $. Here it is required that this should yield
its quantum counter part, the random phase approximation (RPA) \cite
{Vignale}. The weak coupling calculation from the Ornstein-Zernicke
equation is then%
\begin{equation}
c(r)^{(w)}=h^{RPA}\left(  r\right)  -n\int
d\mathbf{r}^{\prime}\left(
c(\left\vert \mathbf{r-r}^{\prime}\right\vert )\right)  ^{(w)}h^{RPA}%
(r^{\prime}).\label{5.7}%
\end{equation}
This has the solution%
\begin{equation}
c(r)^{(w)}=\frac{1}{n}\int\frac{d\mathbf{k}}{\left(  2\pi\right)  ^{3}%
}e^{-i\mathbf{k\cdot r}}\frac{S^{RPA}\left(  k\right)  -1}{S^{RPA}%
(k)}.\label{5.8}%
\end{equation}
Here $S^{RPA}\left(  k\right)  $ is the RPA static structure factor%
\begin{equation}
S^{RPA}\left(  k\right)  =1+n\int d\mathbf{r}e^{i\mathbf{k\cdot r}%
}h^{RPA}(r).\label{5.9}%
\end{equation}
Finally the modified Coulomb potential $\Delta\left(  r\right)  $ in
(\ref{8.1}) becomes
\begin{equation}
\Delta\left(  r\right)
\rightarrow\frac{1}{n}\int\frac{d\mathbf{k}}{\left( 2\pi\right)
^{3}}e^{-i\mathbf{k\cdot r}}\left(  \frac{1}{S^{RPA}(k)}-\frac
{1}{S^{(0)}(k)}\right)  .\label{5.8}
\end{equation}
The definition of $S^{RPA}\left(  k\right)  $ in terms of the RPA
dielectric is given in Appendix \ref{apC}.

\begin{figure}[h]
\centering
\includegraphics[width=80mm]{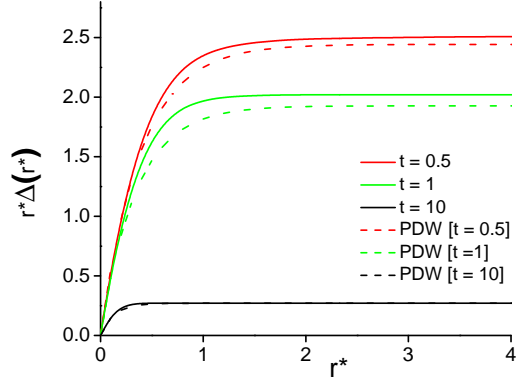}\caption{(color online) Demonstration of crossover for $r^*— \triangle(t, r_s , r^* )$ to Coulomb with effective coupling
constant $\Gamma_{e} (t, r_s )$ given by eq.(\ref {5.12}), for $r_s
= 5$ and $t = 0.5, 1, 10$. Also shown are the corresponding results
for $r^*— \triangle_{PDW} (t, r_s , r^* )$.
 } \label{fig4}
\end{figure}

Several limits of $\Delta\left(  r\right)  =\Delta\left(
t,r_{s},r^{\ast
}\right)  $ are established in Appendix \ref{apC}. For large $r^{\ast}%
=r/r_{0}$ it behaves as $r^{\ast-1}$%
\begin{equation}
\lim_{r^{\ast}}\Delta\left(  t,r_{s},r^{\ast}\right)
\rightarrow\Gamma
_{e}\left(  t,r_{s}\right)  r^{\ast-1},\label{5.11}%
\end{equation}
where $\Gamma_{e}\left(  t,r_{s}\right)  $ is an effective Coulomb
coupling constant
\begin{equation}
\Gamma_{e}\left( t,r_{s}\right)
=\frac{2}{\beta\hbar\omega_{p}\coth\left( \beta\hbar
\omega_{p}/2\right)  }\Gamma,\hspace{0.2in}\Gamma\equiv\frac{\beta q^{2}%
}{r_{0}}.\label{5.12}%
\end{equation}
Here $\omega_{p}=\sqrt{4\pi nq^{2}/m}$ is the plasma frequency. The
dimensionless parameter is $\beta\hbar\omega_{p}=\left(  4/3\right)
\left( 2\sqrt{3}/\pi^{2}\right)  ^{1/3}\sqrt{r_{s}}/t$ so for fixed
$r_{s}$ the high and low temperature limits are
\begin{equation}
\Gamma_{e}\rightarrow\left\{
\begin{array}
[c]{c}%
\Gamma,\hspace{0.2in}\beta\hbar\omega_{p}<<1\\
\left(  \frac{4}{3}r_{s}\right)
^{1/2},\hspace{0.2in}\beta\hbar\omega_{p}>>1
\end{array}
\right.  \label{5.13}%
\end{equation}
This asymptotic Coulomb form is exact and its coefficient follows from the fact that
the RPA incorporates the exact perfect screening sum rule
\cite{Martin99}. It is  illustrated for $r^{\ast}\Delta\left(
t,r_{s},r^{\ast}\right) $ in Figure \ref{fig4} at $r_{s}=5$ for
several values of $t$.

Also shown in this figure are the results from the PDW classical
potential
\begin{equation}
\left(  \beta_{c}\phi_{c}(r)\right)  ^{PDW}=\left(  \beta_{c}\phi
_{c}(r)\right)  ^{(0)}+\Delta^{PDW}\left(  r\right)  .\label{5.14}%
\end{equation}
The Pauli potential $\left(  \beta_{c}\phi_{c}(r)\right)  ^{(0)}$ as
in is the same as in (\ref{8.1}) but its correction
$\Delta^{PDW}\left(  r\right)  $ is
given by the Deutsch regularized Coulomb potential \cite{Deutsch}%
\begin{equation}
\Delta^{PDW}\left(  r\right)  =\beta^{PDW}\frac{q^{2}}{r}\left(
1-e^{-r/\lambda^{PDW}}\right)  ,\hspace{0.2in}\lambda^{PDW}=\left(
\frac{\beta^{PDW}\hbar^{2}}{\pi m}\right)
^{1/2}. \label{5.15}%
\end{equation}
Here $\beta^{PDW}=1/k_{B}T^{PDW}$, where $T^{PDW}\equiv\left(  T^{2}+T_{0}%
^{2}\right)  ^{1/2}$. The only free parameter, $T_{0}=T_{0}\left(
r_{s}\right)  $, is fit by requiring that the classical correlation
energy matches the quantum exchange/correlation energy obtained from
quantum
simulation at $T=0$. The fit given in reference  \cite{DWP00} is%
\begin{equation}
T_{0}\simeq\frac{T_{F}}{a+b\sqrt{r_{s}}+cr_{s}},\label{5.16}%
\end{equation}
with $a=1.594,b=-0.3160,$ and $c=0.0240$. It is seen that the PDW
model is quite similar  to the approximation defined here at
$r_{s}=5$. Greater discrepancies occur for both larger and smaller
$r_{s}$ except at higher temperatures. Further comments on this
comparison are given below.

For $r^{\ast}<<1$, $\Delta\left(  t,r_{s},r^{\ast}\right)  $ approaches a
finite value%
\begin{equation}
\Delta\left(  t,r_{s},0\right)  =\frac{1}{n}\int\frac{d\mathbf{k}}{\left(
2\pi\right)  ^{3}}\left(  \frac{1}{S^{RPA}(k)}-\frac{1}{S^{(0)}(k)}\right)  .
\label{5.16}%
\end{equation}
The integral converges because the static structure factors for
large $k$ approach $1$ as $k^{-4}$ due to quantum effects (cusp
condition \cite {RKB78}). The Coulomb singularity is therefore
removed in the effective classical pair potential. Finally, another
limit obtained in the Appendix \ref {apC} is that for large $r_{s}$
and large $t$ (low
density, high temperature) in which case the Kelbg potential \cite {Kelbg} is recovered%
\begin{equation}
\lim_{t,r_{s}>>1}\Delta\left(  t,r_{s},r^{\ast}\right)  \rightarrow
\Delta_{K}\left(  t,r^{\ast}\right)  =\frac{\Gamma}{r^{\ast}}\left(
1-\exp(2\pi\frac{r_{0}^{2}}{\lambda_{K}^{2}}r^{\ast2})+\sqrt{2}\pi\frac{r_{0}%
}{\lambda_{K}}r^{\ast}(1-\operatorname{erf}(\sqrt{2\pi}\frac{r_{0}}%
{\lambda_{K}}r^{\ast}))\right)  , \label{5.17}%
\end{equation}
with $\lambda_{k}=\lambda/\sqrt{2\pi}$. The Kelbg potential is the
exact weak coupling effective classical potential determined from
the two particle electron - electron density matrix \cite{Kelbg}.
This limit is approached to within $10$ percent at $t=10$ and $1\leq
r_{s}\leq 10$.

\subsubsection{Classical effective temperature, chemical potential}

The approximate temperature and chemical potential equations are obtained in a
similar way%
\begin{equation}
\beta_{c}=\beta_{c}^{(0)}+\left(  \beta_{c}^{RPA}-\beta_{c}^{RPA,(0)}\right)
. \label{5.18}%
\end{equation}%
\begin{equation}
\beta_{c}\mu_{c}=\left(  \beta_{c}\mu_{c}\right)  ^{(0)}+\left(  \beta_{c}%
\mu_{c}\right)  ^{RPA}-\left(  \beta_{c}\mu_{c}\right)  ^{RPA,(0)},
\label{5.19}%
\end{equation}
where $ \beta_{c} ^{(0)}$ \ and  $\left( \beta _{c}\mu_{c}\right)
^{(0)}$ denote the ideal gas results of \cite {DD12}, and
from (\ref{5.1}) and (\ref{5.2})%
\begin{equation}
\beta_{c}^{RPA}=\frac{n\left[  1-\frac{n}{6}\int d\mathbf{r}h^{RPA}%
(r)\mathbf{r}\cdot\nabla\left(  \beta_{c}\phi_{c}(r)\right)  ^{RPA}\right]
}{p^{RPA}}, \label{5.20}%
\end{equation}%
\begin{equation}
\left(  \beta_{c}\mu_{c}\right)  ^{RPA}=\frac{3}{2}\ln\left(  \frac{\beta_{c}^{RPA}%
}{\beta}\right)  +\ln\left(  n\lambda^{3}\right)  ^{RPA}+\frac{1}{2}n\int
d\mathbf{r}h^{RPA}(r)\left(  h^{RPA}(r)+\left(  \beta_{c}\phi_{c}(r)\right)
^{RPA}\right)  . \label{5.21}%
\end{equation}
The RPA results for $p^{RPA}$ and $\left(  n\lambda^{3}\right)  ^{RPA}$ are
computed from the Pade fits of reference \cite{KSK05}.

\begin{figure}[h]
\centering
\includegraphics[width=80mm]{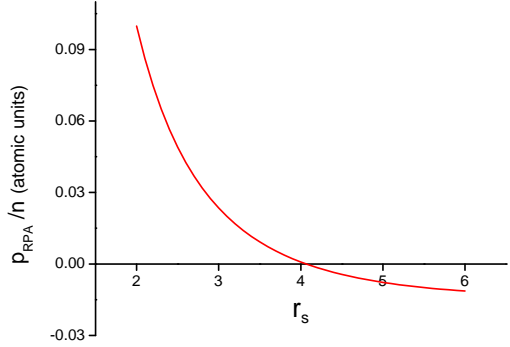}\caption{(color online) Quantum RPA pressure $p_{RPA}$ at $t = 0$ as a function of $r_s$ .
 } \label{fig6}
\end{figure}

\begin{figure}[h]
\centering
\includegraphics[width=80mm]{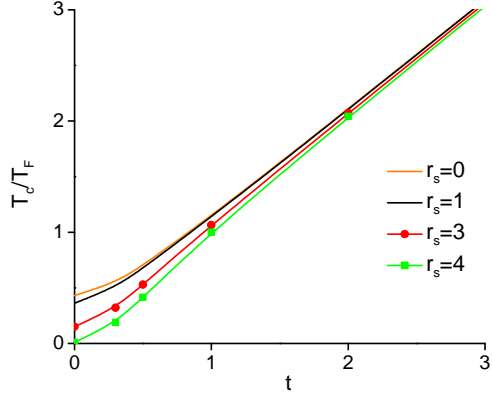}\caption{(color online) Classical reduced temperature $ T_{c} /T_F$ as a function of t for $r_{s} = 0, 1, 3$ and $4$.}
\label{fig7}
\end{figure}

\begin{figure}[h]
\centering
\includegraphics[width=80mm]{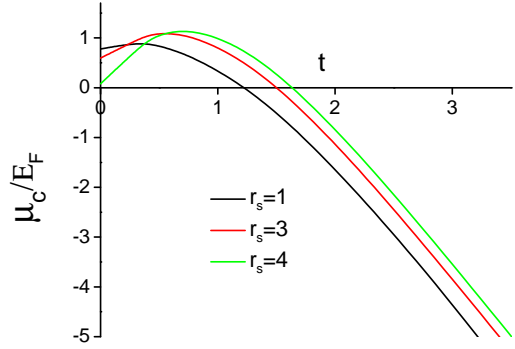}\caption{(color online) Dimensionless classical chemical potential $\mu_c/E_F$ as a function of $t$ for $r_s = 1, 3, 5$.}
\label{fig8}
\end{figure}

A peculiarity of jellium is the possibility for the pressure to
become negative at large $r_{s}$ and small $t$, conditions for which
the equal pressures correspondence condition cannot be imposed. For
real systems, the pressure is positive as follows from the convexity
of the free energy as a function of the volume. This convexity does
not hold for jellium \cite {Baus}. To be more explicit, it is first
noted that the pressure is an increasing function of $t$ so its
minimum value occurs at $t=0$. Figure \ref{fig6} shows the quantum
$p^{RPA}$ at $t=0$ as a function on $r_{s}$. It is seen that
$p^{RPA}\left( t=0\right)  $ vanishes for $r_{s}\simeq4$, and
becomes negative for larger $r_{s}$. Thus for $r_{s}\gtrsim4$ the
pressure $p^{RPA}\left( t\right)  $ vanishes at some temperature
$t_{0}\left(  r_{s}\right) $. Then from (\ref{5.18}) and
(\ref{5.20}), the effective classical temperature vanishes at
$t_{0}$
\begin{equation}
T_{c}\left(  t_{0}\left(  r_{s}\right)  \right)  =\frac{p^{RPA}\left(
t_{0}\right)  }{n\left[  1-\frac{n}{6}\int d\mathbf{r}h^{RPA}(r)\mathbf{r}%
\cdot\nabla\left(  \beta_{c}\phi_{c}(r)\right)  ^{RPA}\right]  }=0.
\label{5.21a}%
\end{equation}
For $t<t_{0}\left(  r_{s}\right)$, $p^{RPA}\left(  t\right)  <0$.
However, it is found that the denominator of (\ref{5.21a}) remains
positive. Since the classical temperature must be positive this
indicates that the equivalence condition, $p_{cl}=p$, can no longer
be realized. Therefore for jellium, this one of the
equivalence conditions should be replaced by a different condition
(e.g., equivalence of internal energies). Instead, the analysis here
is restricted to $t>$ $t_{0}\left(  r_{s}\right)  $ to assure
positive pressure. Figure \ref{fig7} shows $t_{c}=T_{c}/T_{F}$ as a
function of $t$ calculated from (\ref{5.18}) for $r_{s}=0,1,3,4,$ and
$5$. Figure \ref{fig8} shows the corresponding results for
$\mu_{c}/E_F$ calculated from (\ref{5.19}).

\subsection{Radial distribution function and thermodynamics}

With the parameters of the effective classical system determined
approximately above they can be used in an accurate classical
many-body method to combine the quantum properties of these
parameters with classical strong coupling effects (e.g., molecular
dynamics simulation). This is illustrated here by using the full HNC
integral equations (\ref{5.3}) and (\ref{5.4}) specialized to
calculate $g(r)$ for jellium
\begin{equation}
\ln g(r)=-\beta_{c}\phi_{c}(r)+h\left(  r\right)  -c\left(  r\right)
,\hspace{0.2in}c\left(  r\right)  =h\left(  r\right)  -n\int d\mathbf{r}%
^{\prime}c(\left\vert \mathbf{r-r}^{\prime}\right\vert )h\left(  r^{\prime
}\right)  . \label{5.22}%
\end{equation}
Of course, these equations are also those used to define $\beta_{c}\phi
_{c}(r)$ so the analysis would seem to be circular. However, the approach has
been to use an approximation to the HNC equations to determine $\beta_{c}%
\phi_{c}$ (here the weak coupling RPA limit) and then to "bootstrap" this
information to solve the full HNC equations for a $g(r)$ that goes beyond the
input $g^{RPA}(r)$ to include classical strong coupling. One manifest
improvement obtained in this way is positivity of $g(r)$, already noted in
\cite{DWP00}. In contrast, $g^{RPA}(r)$ becomes negative for small $r$ at
sufficiently large $r_{s}$.

\begin{figure}[h]
\centering
\includegraphics[width=80mm]{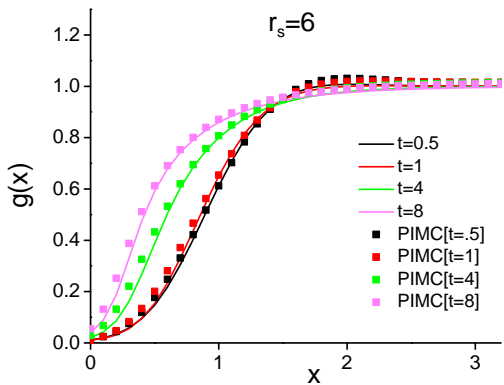}\caption{(color online) Radial distribution function $g(r^*)$ for $r_s = 6$ at $t = 0.5, 1, 4, 8$.
Also shown are the results of PIMC. }\label{fig9}
\end{figure}

\begin{figure}[h]
\centering
\includegraphics[width=80mm]{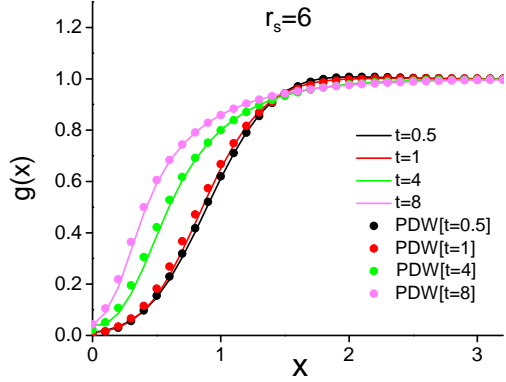}\caption{(color online) Radial distribution function $g(r^*)$ for $r_s = 6$ at $t = 0.5, 1, 4, 8$. Also shown are
the results of PDW.}\label{fig10}
\end{figure}

The determination of $g(r)$ from (\ref{5.22}) is straightforward
using the method described in reference \cite{Ng}. Note that these
equations do not use the equal pressure condition nor the value of
$\beta_{c}$. Hence they do not have the restriction to positive
pressures and the associated restriction on $r_{s}.$ The results are
shown in Figure \ref{fig9} for the case of $r_{s}=6$ at $t=0.5,1,4$
and $8$. Also shown are the results from recent restricted PIMC
\cite{Brown}. The agreement is quite good. Figure \ref{fig10} shows
the same conditions as Figure \ref{fig9} for comparison with the
classical map of PDW. The agreement is remarkable given that the
forms and origins of the effective classical parameters is so
different. This agreement between the predictions here, PIMC, and
PDW extends to other state conditions as well, except for small $t$
and very large $r_s$.

Other theoretical models for $g(r)$ are based on the same dielectric
formalism of the RPA, but including "local field" corrections. One
of the earliest was the self-consistent STLS model \cite{STLS70} at
$T=0$, later generalized to finite temperature $T$ \ by Tanaka and
Ichimaru (TI) \cite{Tanaka86}.  The discrepancies (not shown) are
largest at lower $t$ and most noticeably at small distances where TI
becomes negative. The RPA results are significantly more negative in
this range. Both RPA and its improved TI overestimate the size of
the electron correlation hole \cite{Vignale} at larger values for
$r_{s}$.

\begin{figure}[h]
\centering
\includegraphics[width=80mm]{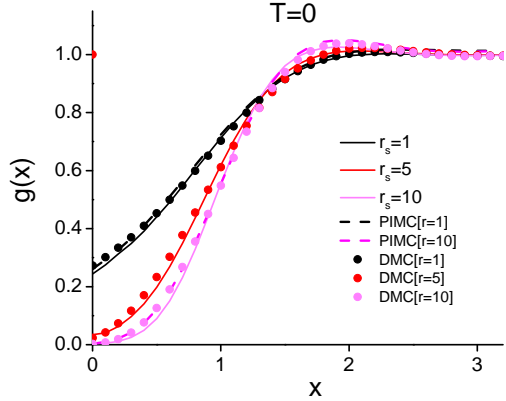}\caption{ Radial distribution function $g(r^*)$ for $t = 0$ at $r_s = 1, 5, 10$. Also shown are
results from PIMC and diffusion Monte Carlo. The PIMC and diffusion Monte Carlo plots are indistinguishable.
}\label{fig11}
\end{figure}

The PDW $g(r)$ is in good agreement with diffusion MC data at $t=0$
\cite{DWP00} for $r_{s}=1,5,10$. Although the single parameter
$T_{0}$ is fixed by fitting the $t=0$ exchange/correlation energy
from MC data it is nevertheless impressive that this provides good
results for $g(r)$ across a range both $r_{s}$ and $r$. Figure
\ref{fig11} shows a comparison of the results of the present
analysis with the same $T=0$ diffusion MC data \cite {Ortiz}, and
also the recent PIMC for $T=0.065$ at $r_{s}=1,10$. The good
agreement is quite surprising since there is no MC parametrization
in the present analysis and all quantum input is via the RPA and
ideal gas exchange. However, it is recalled that the RPA preserves
the exact quantum mechanics of the perfect screening sum rule that
governs the cross over to exact large $r^*$ Coulomb limit. This is
discussed further in the last section.

\subsubsection{Thermodynamics}

\begin{figure}[h]
\centering
\includegraphics[width=80mm]{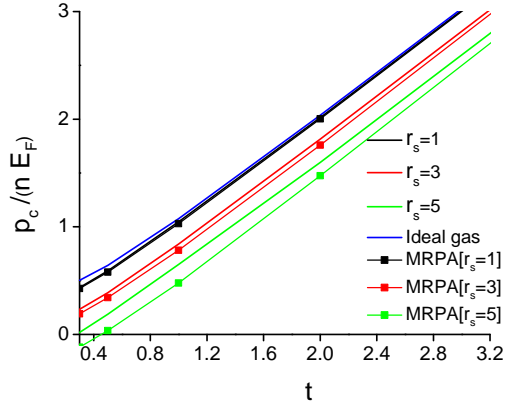}\caption{(color online) Dimensionless classical pressure $p_c/(nE_F)$ as a function of t for $r_s = 1, 3, 5$.
Also shown are the corresponding modified RPA results.}\label{fig12}
\end{figure}

The predicted pressure, $p_c$ in atomic units, for the effective classical system is obtained from%
\begin{equation}
\frac{\beta_{c}p_{c}}{n_{c}}=1-\frac{1}{6}n\int d\mathbf{r}h(r)\mathbf{r}%
\cdot\nabla\beta_{c}\phi_{c}(r).\label{5.23}%
\end{equation}
and the effective temperature (\ref{5.20}).
Figure \ref{fig12} shows this as a function of $t$ for $r_{s}=1,3,4$
and $5$. Also shown are the corresponding results for modified RPA (using the
fits from reference \cite{PDW1984}).

\section{Application to Charges in a Harmonic Trap}

\label{sec6}As a final application here consider $\overline{N}$
charges localized within a harmonic trap. The Hamiltonian is
\begin{equation}
H-\mu N=\sum_{i=1}^{N}\frac{p_{i}^{2}}{2m}+\frac{1}{2}\sum_{i\neq j}^{N}%
\frac{q^{2}}{\left\vert \mathbf{r}_{i}-\mathbf{r}_{j}\right\vert }-\int
d\mathbf{r}\mu(\mathbf{r})\widehat{n}(\mathbf{r}), \label{6.1}%
\end{equation}
with the local chemical potential given explicitly as%
\begin{equation}
\mu(\mathbf{r})=\mu-\frac{1}{2}m\omega^{2}r^{2}. \label{6.2}%
\end{equation}
The constant $\mu$ determines the average number of charges $\overline{N}$. As
a consequence of the harmonic potential the equilibrium average density
profile for the charges is non-uniform and depends only on the radial
coordinate%
\begin{equation}
n(r)=\Omega^{-1}\sum_{N=0}^{\infty}N\int d\mathbf{r}_{2}..d\mathbf{r}%
_{N}\left\langle \mathbf{r}_{1}..\mathbf{r}_{N}\right\vert
e^{-\beta\left(  H-\mu N\right)  }\left\vert
\mathbf{r}_{1}..\mathbf{r}_{N}\right\rangle ,
\label{6.3}%
\end{equation}
where $\left\langle \mathbf{r}_{1}..\mathbf{r}_{N}\right\vert
X\left\vert \mathbf{r}_{1}..\mathbf{r}_{N}\right\rangle $ is the $N$
particle diagonal anti-symmetric matrix element in coordinate
representation, and $\Omega$ is the grand potential. The density
profile in the classical limit has been studied in detail, via
simulation and theory \cite{Wrighton09}. In that case the
dimensionless form depends on $r_{s}$ and $t$ only through the
Coulomb coupling constant $\Gamma=\beta q^{2}/r_{0}=\left(
4/3\right)  \left( 2/3\pi^{2}\right)  ^{1/3}r_{s}/t$. For
sufficiently large $\Gamma$ the formation of shell structure is
observed in $n(r)$. The objective now is to exploit this effective
classical description to explore the effects of quantum diffraction
and exchange via the proposed effective classical system. Only a
preliminary investigation of new mechanisms for shell structure is
described here, with a more complete discussion to be given
elsewhere.

The basis for the study is the HNC description for the inhomogeneous
case, Eq. (37) of reference \cite {DD12}
\begin{equation}
\ln\left(  n\left(  \mathbf{r}\right)  \lambda_{c}^{3}\right)  =\beta_{c}%
\mu_{c}(\mathbf{r})+\int d\mathbf{r}^{\prime}c^{(2)}(\mathbf{r,r}^{\prime}\mid
n)n\left(  \mathbf{r}^{\prime}\right)  . \label{6.5}%
\end{equation}
The classical studies of reference \cite{Wrighton09} made a further
approximation to this expression, replacing the correlations for the
inhomogeneous system $c^{(2)}(\mathbf{r,r}^{\prime\prime}\mid n)$ by
those for a corresponding uniform one component plasmas (OCP or
classical jellium), $c^{(2)}(\mathbf{r,r}^{\prime\prime}\mid
n)\rightarrow c(\left\vert \mathbf{r-r}^{\prime\prime}\right\vert
,n)$. The results based on this approximation are found to be quite
accurate except at very strong coupling. A partial theoretical basis
has been given \cite{Wrighton12}. This approximation will be made
here as well.

It is convenient to rewrite (\ref{6.5}) in a Boltzmann form with an effective
potential $U(\mathbf{r})$ defined by
\begin{equation}
n\left(  \mathbf{r}\right)  =\overline{N}\frac{e^{-\beta_{c}U(\mathbf{r})}%
}{\int d\mathbf{r}^{\prime}e^{-\beta_{c}U(\mathbf{r}^{\prime})}}, \label{6.6}%
\end{equation}
so that (\ref{6.5}) becomes
\begin{equation}
\beta_{c}U(\mathbf{r})=-\beta_{c}\left(  \mu_{c}(\mathbf{r})-\mu_{c}\right)
-\frac{\overline{N}}{\int d\mathbf{r}^{\prime}e^{-\beta_{c}U(\mathbf{r}%
^{\prime})}}\int d\mathbf{r}^{\prime}e^{-\beta_{c}U(\mathbf{r}^{\prime}%
)}c(\left\vert \mathbf{r-r}^{\prime}\right\vert ,n). \label{6.7}%
\end{equation}
Practical application of this result requires specification of the direct
correlation function $c(r,n)$ for jellium and the classical local chemical
potential $\mu_{c}(\mathbf{r})-\mu_{c}$. The former is determined from the
equivalent classical calculation described in the previous section. The latter
is the effective classical trap potential corresponding to the actual quantum
harmonic trap. It's approximate determination is described in the next subsection.

The total number of particles appears explicitly. To introduce the density, it
is necessary to assign a volume for the system. This can be defined as the
volume of a sphere with radius $R_{0}$ corresponding to a particle at the
greatest distance from the center. At equilibrium the average density can be
taken to be spherically symmetric so that the total average force on that
particle is%
\begin{equation}
\frac{\overline{N}q^{2}}{R_{0}^{2}}-m\omega^{2}R_{0}=0,\hspace{0.25in}%
\Rightarrow R_{0}^{3}=\overline{N}\frac{q^{2}}{m\omega^{2}}. \label{6.7a}%
\end{equation}
This gives the average density to be
\begin{equation}
\overline{n}\equiv\frac{3\overline{N}}{4\pi R_{0}^{3}}=\frac{3m\omega^{2}%
}{4\pi q^{2}}. \label{3.7b}%
\end{equation}
In this way, the trap parameter $m\omega^{2}/q^{2}$ is specified in terms of
the density.

\subsection{Approximate form for $\mu_{c}(\mathbf{r})$}

\begin{figure}[h]
\centering
\includegraphics[width=80mm]{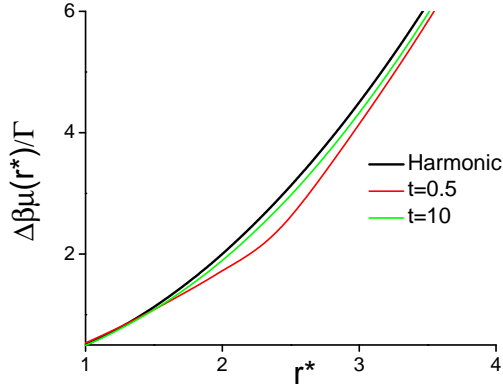}\caption{(color online) Effective classical trap potential $\triangle\beta_c
\mu_c(\mathbf{r}^*)/\Gamma=(\beta_c \mu_c(\mathbf{r}^*)-\beta_c
\mu_c(\mathbf{0}))/\Gamma$ as a function of $r^*$ for $t = 0.5, 10$. Also shown is the harmonic potential. }\label{fig13}
\end{figure}

Without quantum effects $\mu_{c}-\mu_{c}(\mathbf{r})$ is just the harmonic
potential of (\ref{6.2}). Modifications for the effective classical form occur
due to both diffraction and exchange effects. The exchange effects are
dominated by those for the ideal Fermi gas in a harmonic trap. As a first
approximation here, $\mu_{c}(\mathbf{r})$ is replaced by that for an ideal gas
of $\overline{N}$ Fermions in a harmonic trap. This is the inhomogeneous ideal
gas considered in Section IV in \cite{DD12}. The local chemical potential is therefore
obtained from (\ref{6.5}) specialized to an ideal gas
\begin{equation}
(\beta_{c}\mu_{c})^{(0)}(\mathbf{r})=\ln(n^{(0)}(\mathbf{r})\lambda_{c}%
^{3})-\int d\mathbf{r}^{\prime}c^{(0)}(\left\vert \mathbf{r}-\mathbf{r}%
^{\prime}\right\vert )n^{(0)}(\mathbf{r}^{\prime}). \label{6.8}%
\end{equation}%
Then the effective potential $U(\mathbf{r})$ becomes
\begin{equation}
\beta_{c}U(\mathbf{r})=-\ln(n^{(0)}(\mathbf{r})\lambda_{c}^{3})-\int
d\mathbf{r}^{\prime}c^{(0)}(\left\vert \mathbf{r}-\mathbf{r}^{\prime
}\right\vert )n^{(0)}(\mathbf{r}^{\prime})+\beta_{c}\mu_{c}
-\overline{N}\int
d\mathbf{r}^{\prime}\frac{e^{-\beta_{c}U(\mathbf{r}^{\prime})}}{\int
d\mathbf{r}^{\prime\prime}e^{-\beta_{c}U(\mathbf{r}^{\prime\prime})}%
}c(\left\vert \mathbf{r-r}^{\prime}\right\vert ,n). \label{6.8a}
\end{equation}
The direct correlation function for the uniform ideal Fermi gas,
$c^{(0)}(r,n)$, is
known from the results of \cite{DD12}. Furthermore, $n^{(0)}%
(\mathbf{r})$ has an explicit form in the local density
approximation of Appendix B, reference \cite {DD12}
\begin{equation}
n^{(0)}(\mathbf{r})=\frac{2}{\lambda^{3}}f_{3/2}\left(  e^{\left(  \beta
\mu^{(0)}-\frac{1}{2}\Gamma r^{\ast2}\right)  }\right)  . \label{6.9}%
\end{equation}
where $\Gamma$ is the same coupling constant as in eqn (\ref{5.12}). The Fermi function $f_{3/2}\left(  x\right)  $ is given by
\begin{equation}
f_{3/2}(z)=\frac{4}{\sqrt{\pi}}\int_{0}^{\infty}dxx^{2}\left(  z^{-1}e^{x^{2}%
}+1\right)  ^{-1}. \label{6.9a}%
\end{equation}
The constant chemical potential $\mu^{(0)}$ is determined in terms
of $t,r_{s}$ by the condition that the average number of particles
is
$\overline{N}$%
\begin{equation}
\overline{N}=2\left(  \frac{r_{0}}{\lambda}\right)  ^{3}\int d\mathbf{r}%
^{\ast}f_{3/2}\left(  e^{\beta\left(  \mu^{(0)}-\frac{1}{2}\Gamma r^{\ast
2}\right)  }\right)  . \label{6.10}%
\end{equation}
Hence $n^{(0)}(\mathbf{r})r_{0}^{3}$ is also given by (\ref{6.9}) in terms of
$t,r_{s}.$ Figure \ref{fig13} shows $\beta_{c}\phi_{cext}(\mathbf{r}%
)\equiv(\beta_{c}\mu_{c})^{(0)}(\mathbf{r})-(\beta_{c}\mu_{c})^{(0)}%
(\mathbf{0})$ obtained in this way for $r_{s}=5$ and $t=0.5,1,5,$ and $10$.
The classical trap potential is harmonic at large $r$, but there are
significant deviations at the lower temperatures for $2\lesssim r\lesssim3$.

\subsection{Quantum effects on the mean field density profile}

\begin{figure}[h]
\centering
\includegraphics[width=80mm]{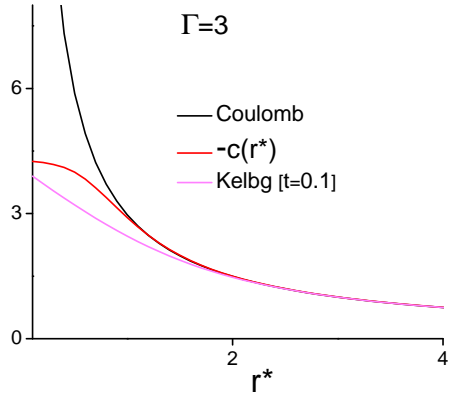}\caption{(color online) Comparison of $-c(r^*)$ and $−βV_K (r^*)$ at $t = 0.1, 0.27$ both
corresponding to $\Gamma = 3$. Also shown is the Coulomb limit
$−\beta q^2 /r$. }\label{fig14}
\end{figure}

The density profile can now be determined from (\ref{6.7}) where the effective
potential of (\ref{6.8}) becomes
\begin{equation}
\beta_{c}U(\mathbf{r})=\beta_{c}\phi_{c,ext}(\mathbf{r})-\overline{N}\int
d\mathbf{r}^{\prime}\frac{e^{-\beta_{c}U(\mathbf{r}^{\prime})}}{\int
d\mathbf{r}^{\prime\prime}e^{-\beta_{c}U(\mathbf{r}^{\prime\prime})}%
}c(\left\vert \mathbf{r-r}^{\prime}\right\vert ,n).\label{6.11}%
\end{equation}
(the potential $\beta_{c}U(\mathbf{r})$\ has been shifted by a constant to
simplify the result). The effective classical external trap potential is%
\begin{equation}
\beta_{c}\phi_{c,ext}(\mathbf{r})=-\ln(n^{(0)}(\mathbf{r})\lambda^{3})+\int
d\mathbf{r}^{\prime}c^{(0)}(\left\vert \mathbf{r}-\mathbf{r}^{\prime
}\right\vert )n^{(0)}(\mathbf{r}^{\prime})\label{6.12}%
\end{equation}
Quantum effects result from the deviation of $\beta_{c}\phi_{c,ext}%
(\mathbf{r})$ from the given harmonic potential, and the deviation of $c(r,n)$
from its classical \ OCP form. In this section these two sources are isolated
to explore the possibility of new origins of shell structure. To do so only
the mean field limits of (\ref{6.11}) and (\ref{6.12}) are explored here. The
mean field limit is defined by $c(r,n)\rightarrow-\beta_{c}\phi_{c}(r)$ and
$c^{(0)}(r,n)\rightarrow-\left(  \beta_{c}\phi_{c}(r)\right)  ^{(0)}$so
(\ref{6.11}) and (\ref{6.12}) become%
\begin{equation}
\beta_{c}U(\mathbf{r})\rightarrow\beta_{c}\phi_{c,ext}(\mathbf{r}%
)+\overline{N}\int d\mathbf{r}^{\prime}\frac{e^{-\beta_{c}U(\mathbf{r}%
^{\prime})}}{\int d\mathbf{r}^{\prime}e^{-\beta_{c}U(\mathbf{r}^{\prime})}%
}\left[  \left(  \beta_{c}\phi_{c}(\left\vert \mathbf{r-r}^{\prime}\right\vert
)\right)  ^{(0)}+\Delta\left(  \left\vert \mathbf{r-r}^{\prime}\right\vert
\right)  \right]  ,\label{6.13}%
\end{equation}%
\begin{equation}
\beta_{c}\phi_{c,ext}(\mathbf{r})\rightarrow-\ln(n^{(0)}(\mathbf{r}%
)\lambda^{3})-\int d\mathbf{r}^{\prime}\left(  \beta_{c}\phi_{c}(\left\vert
\mathbf{r-r}^{\prime}\right\vert )\right)  ^{(0)}n^{(0)}(\mathbf{r}^{\prime
}).\label{6.14}%
\end{equation}
The classical mean free limit corresponds to $\left(  \beta_{c}\phi
_{c}(r)\right)  ^{(0)}=0$ and $\Delta\left(  r\right)  =\beta
q^{2}/r$. There is no shell structure in this limit, even at very
strong coupling. Instead, shell structure arises due to sufficiently
large Coulomb coupling, such that $c(r,n)$ differs from $-\beta
q^{2}/r$ inside the correlation length $r_{0}$ and the Coulomb
singularity at $r=0$ is removed. This is shown in Figure \ref{fig14}
for $\Gamma=3$.

\subsubsection{Diffraction effects}

\begin{figure}[h]
\centering
\includegraphics[width=80mm]{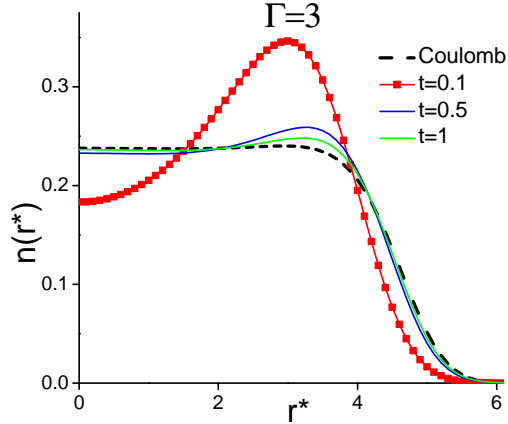}\caption{(color online) Diffraction mean field approximate density profile for $\Gamma=3$ and $t = 0.1, 0.5, 1$.
}\label{fig15}
\end{figure}

To explore the effects of diffraction only in (\ref{6.13}) and (\ref{6.14}),
the contributions from exchange are set to zero, i.e. $\left(  \beta_{c}%
\phi_{c}(r)\right)  ^{(0)}\rightarrow0,\beta_{c}\phi_{c,ext}(\mathbf{r}%
)\rightarrow m\omega^{2}r^{2}/2,$ and $\Delta\left(  r\right)  \rightarrow
\Delta_{K}\left(  r\right)  =$ Kelbg, eq. (\ref{5.17}). Then (\ref{6.13}) becomes
\begin{equation}
\beta_{c}U(\mathbf{r})\rightarrow\beta_{c}\frac{1}{2}m\omega^{2}%
r^{2}+\overline{N}\int d\mathbf{r}^{\prime}\frac{e^{-\beta_{c}U(\mathbf{r}%
^{\prime})}}{\int d\mathbf{r}^{\prime}e^{-\beta_{c}U(\mathbf{r}^{\prime})}%
}\beta\Delta_{K}\left(  \left\vert \mathbf{r-r}^{\prime}\right\vert \right)
,\label{6.15}%
\end{equation}
This has the same form as the classical limit, except with the
Coulomb potential replaced by the Kelbg form. The latter differs
from Coulomb at short distances, for which it is finite at $r=0$.
Thus diffraction (without Coulomb correlations) leads to the same
qualitative physical effects as classical Coulomb correlations. This
is illustrated in Figure \ref{fig14} where the Kelbg potential is
evaluated at $r_s =0.042$, $t=0.1$ and at $r_s =0.11$, $t=0.27$ both
corresponding to $\Gamma=3$. For this reason it can be expected that
the quantum diffraction mean field approximation can give rise to
shell structure not present in the corresponding classical case.
This is shown in Figure \ref{fig15} for $\Gamma=3$ and
$t=0.1,0.5,1,2$. A clear shell formation occurs at the two lowest
temperatures, for which the diffraction regularization of
$\Delta_{K}\left( \mathbf{0}\right)  $ is greatest.

\subsubsection{Exchange effects}

\begin{figure}[h]
\centering
\includegraphics[width=80mm]{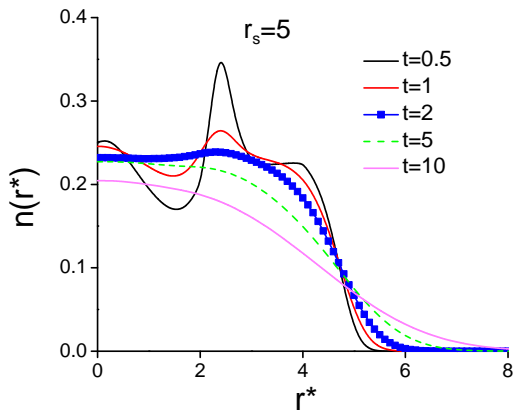}\caption{(color online) Exchange mean field approximate density profile for $r_s = 5$ and $t = 0.5, 1, 2, 5,10$.
}\label{fig16}
\end{figure}

Now return to the mean field form (\ref{6.13}) and neglect all diffraction
effects to study the effects of exchange only
\begin{equation}
\beta_{c}U(\mathbf{r})\rightarrow\beta_{c}\phi_{c,ext}(\mathbf{r}%
)+\overline{N}\int d\mathbf{r}^{\prime}\frac{e^{-\beta_{c}U(\mathbf{r}%
^{\prime})}}{\int d\mathbf{r}^{\prime}e^{-\beta_{c}U(\mathbf{r}^{\prime})}%
}\left[  \left(  \beta_{c}\phi_{c}(\left\vert \mathbf{r-r}^{\prime}\right\vert
)\right)  ^{(0)}+\beta q^{2}\left\vert \mathbf{r-r}^{\prime}\right\vert
^{-1}\right]  ,\label{6.16}%
\end{equation}
This differs from the classical form by the addition of the exchange Pauli
potential to the Coulomb potential, and by the modifications of the harmonic
trap form in $\phi_{c,ext}(\mathbf{r})$, Figure \ref{fig13}.  If the latter
are neglected it is expected that no shell structure will appear, since the
Pauli plus Coulomb potential is still singular at short range and behaves as
the classical mean field limit. The quantitative changes in $\phi
_{c,ext}(\mathbf{r})$ shown in Figure \ref{fig13} are modest for $t\geq2$, but
there is a qualitative change in shape for $2<$ $r^{\ast}<3$ at lower
temperatures. The corresponding density profiles from (\ref{6.16}) are shown
in Figure \ref{fig16} at $r_{s}=5$. Shell formation in this range is clearly
seen for $t=1$ and $0.5$.

\section{Discussion}
\label{sec7}There are powerful many body methods within classical
equilibrium statistical mechanics that do not apply directly to
quantum systems. Examples are molecular dynamics simulation and
liquid state theory. To bridge this gap a method to define an
equivalent classical system for the thermodynamics and structure of
a given quantum system has been developed \cite{DD12}. The objective
here has been to demonstrate and test that approach with
applications to two quite different quantum systems, the bulk
uniform electron gas and charges confined by a harmonic trap. In the
first case, a simple representation for the pair potential
incorporating both ideal gas exchange and RPA correlations was used
within a strong coupling classical theory - the HNC. Essential
properties such as positivity are assured by the classical
formalism, in contrast to early quantum theories with mean field
corrections to RPA. Good agreement with diffusion Monte Carlo
simulation over a range of densities at $t=0$. Initial comparisons
with recent results at finite $t$ \cite{Brown} also show good
agreement. A more extensive comparison will be discussed elsewhere.

The good agreement at $t=0$ is more than might be expected from the
RPA input for the classical pair potential. The PDW model also has
similar agreement but is parameterized by exchange correlation
simulation data at $t=0$ so agreement is less surprising. One
possible explanation for the results here is the preservation of the
exact perfect screening sum rule. This assures that the effective
potential in the approximation used here has the exact Coulomb tail
for large $r$, (\ref{5.11}) and (\ref{5.12}). Elsewhere, a simple
analytic model is constructed which incorporates this asymptotic
property. Its
comparison with Monte Carlo simulation data for $g(r)$ has accuracy comparable to that of the results presented here. %

The second application here was to charges confined by a harmonic
trap. This is an interesting test system for strong correlations
since classically this is reflected in the formation of radial shell
structure. Here, only the mean field theory (no Coulomb
correlations) was considered as a means to explore the possibility
of purely quantum mechanisms for shell structure. It was found that
diffraction effects, which regularize the Coulomb potential at $r=0$
mimic classical Coulomb correlations and does lead to shell
structure. In addition, in the absence of both diffraction effects
and Coulomb correlations, the changes in the effective confining
potential due to exchange effects also can lead to shell structure.
Elsewhere, a detailed application of this effective classical system
will be described with all mechanisms for shell structure active.

\section{Acknowledgements}

This research has been supported by NSF/DOE Partnership in Basic Plasma
Science and Engineering award DE-FG02-07ER54946 and by US DOE Grant DE-SC0002139.

\appendix

\section{RPA for the Uniform Electron Gas}

\label{apC} In this Appendix the RPA is defined and it is shown that
the modified Coulomb potential reduces to the Kelbg potential for
weak coupling and weak degeneracy. More generally, the exact large
$r$ dependence of this potential is evaluated.

The static structure factor is related to the Fourier transform of
the hole
function by%
\begin{equation}
S(k)=1+n\int d\mathbf{r}e^{i\mathbf{k\cdot r}}h(r). \label{a.1}%
\end{equation}
As a density fluctuation it is also related to the dynamic response
function
or complex dielectric function ${\epsilon(\omega,k)}$ \cite{Vignale}:%
\begin{equation}
S\left(  k\right)
=-\frac{\hbar}{\pi}\frac{1}{\widetilde{V}(k)}\int_{-\infty
}^{\infty}d\omega\left(  1-e^{-\beta\hbar\omega}\right)  ^{-1}%
\operatorname{Im}\epsilon^{-1}(k,\omega), \label{a.2}%
\end{equation}
where $\widetilde{V}(k)=4\pi q^{2}/k^{2}$ is the Fourier transformed
Coulomb
potential. In the random phase approximation the dielectric function is%
\begin{equation}
\epsilon^{RPA}(k,\omega)=1-\widetilde{V}(k)\chi^{(0)}(\omega,k), \label{a.3}%
\end{equation}
and $\chi^{(0)}\left(  \mathbf{k},\omega\right)  $ is the response
function for the ideal Fermi gas
\begin{equation}
\chi^{(0)}\left(  \mathbf{k},\omega\right)  \equiv\frac{\left(
2s+1\right)
}{n}\lim_{\eta\rightarrow0^{+}}\int\frac{d\mathbf{k}_{1}}{\left(
2\pi\right) ^{3}}\frac{n\left(  \epsilon_{\left\vert
\mathbf{k-k}_{1}\right\vert }\right) -n\left(
\epsilon_{k_{1}}\right)  }{\hbar\omega+i\eta+e_{k_{1}}-e_{\left\vert
\mathbf{k-k}_{1}\right\vert
}},\hspace{0.2in}e_{k}=\frac{\hbar^{2}k^{2}}{2m}
\label{a.4}%
\end{equation}

\subsection{Kelbg limit}

For weak coupling $\epsilon^{-1}(k,\omega)$ can be expanded to
quadratic order
in $\widetilde{V}(k)$ to get%
\begin{equation}
S\left(  k\right)  \rightarrow S^{(0)}\left(  k\right)
-2\frac{\hbar}{\pi}\widetilde
{V}(k)\int_{-\infty}^{\infty}d\omega\left(
1-e^{-\beta\hbar\omega}\right) ^{-1}\left(
\operatorname{Im}\chi^{(0)}(\omega,k)\right)  \left(
\operatorname{Re}\chi^{(0)}(\omega,k)\right)  .\label{a.5}%
\end{equation}
The real and imaginary parts of $\chi^{(0)}(\omega,k)$ are%
\begin{equation}
\operatorname{Re}\chi^{(0)}\left(  \mathbf{k},\omega\right)
=-\frac{\left(
2s+1\right)  }{n\lambda^{3}}\beta\frac{1}{4\sqrt{\pi}\kappa}\mathcal{P}%
\int_{-\infty}^{\infty}dx\ln\left(  1+ze^{-x^{2}}\right)  \left(  \frac{1}%
{\nu+\kappa-x}-\frac{1}{\nu-\kappa-x}\right)  \label{a.6}%
\end{equation}%
\begin{equation}
\operatorname{Im}\ \chi^{(0)}\left(  \mathbf{k},\omega\right)
=\frac{\left( 2s+1\right)
}{n\lambda^{3}}\beta\frac{\sqrt{\pi}}{4\kappa}\ln\left(
\frac{1+ze^{-\left(  \nu+\kappa\right)  ^{2}}}{1+ze^{-\left(  \nu
-\kappa\right)  ^{2}}}\right)  .\label{a.7}%
\end{equation}
The dimensionless variables $\kappa$ and $\nu$ are
\begin{equation}
\kappa\mathbf{=}\frac{k\lambda}{4\sqrt{\pi}},\hspace{0.25in}\nu=\frac
{\beta\hbar\omega}{4\kappa},\hspace{0.25in}\lambda=\left(
\frac{2\pi
\beta\hbar^{2}}{m}\right)  ^{1/2}.\label{a.8}%
\end{equation}

Next consider the additional limit of weak degeneracy. This is
implemented by
an expansion in $z$%
\begin{equation}
S^{(0)}\left(  k\right)  \rightarrow1+\mathcal{O}(z),\hspace{0.2in}%
n\lambda^{3}\rightarrow\left(  2s+1\right)  z,\label{a.9}%
\end{equation}%
\begin{equation}
\operatorname{Re}\chi^{(0)}\left(  \mathbf{k},\omega\right)
\rightarrow
\frac{\beta}{4\kappa}\left(  g(\nu+\kappa)-g(\nu-\kappa)\right)  \label{a.10}%
\end{equation}%
\begin{equation}
\operatorname{Im}\ \chi^{(0)}\left(  \mathbf{k},\omega\right)
\rightarrow
\beta\frac{\sqrt{\pi}}{4\kappa}\left(  e^{-\left(  \nu+\kappa\right)  ^{2}%
}-e^{-\left(  \nu-\kappa\right)  ^{2}}\right)  .\label{a.11}%
\end{equation}
with%
\begin{equation}
g\left(  y\right)
=-\frac{1}{\sqrt{\pi}}\mathcal{P}\int_{-\infty}^{\infty
}dxe^{-x^{2}}\frac{1}{y-x}=-2e^{-\nu^{2}}\int_{0}^{\nu}dxe^{x^{2}%
}.\label{a.12}%
\end{equation}
The RPA structure factor becomes%
\begin{align}
S\left(  k\right)    &
\rightarrow1+\beta\widetilde{V}(k)\frac{1}{2\kappa
\sqrt{\pi}}\int_{-\infty}^{\infty}d\nu e^{-\left(  \nu-\kappa\right)  ^{2}%
}\left(  g(\nu+\kappa)-g(\nu-\kappa)\right)  \nonumber\\
& =1+\beta\widetilde{V}(k)\frac{1}{2\kappa\sqrt{2}}g(\sqrt{2}\kappa
)\label{a.13}%
\end{align}

The modified Coulomb potential at weak coupling and weak degeneracy becomes%
\begin{align}
\Delta\left(  r\right)    &
=\frac{1}{n}\int\frac{d\mathbf{k}}{\left( 2\pi\right)
^{3}}e^{-i\mathbf{k\cdot r}}\left(  \frac{1}{S^{RPA}(k)}-\frac
{1}{S^{(0)}(k)}\right)  \nonumber\\
& \rightarrow-\frac{1}{n}\int\frac{d\mathbf{k}}{\left(  2\pi\right)  ^{3}%
}e^{-i\mathbf{k\cdot r}}\beta\widetilde{V}(k)\frac{1}{2\kappa\sqrt{2}}%
g(\sqrt{2}\kappa)\nonumber\\
& =\beta V_{K}(r)\label{a.14}%
\end{align}
This is the Kelbg potential of eq. (\ref{5.17}).

\subsection{Large $r$ limit}

The large $r$ behavior of $\Delta\left(  r\right)  $ is governed by
the small
$k$ behavior of $S^{RPA}(k)$%
\begin{equation}
S^{RPA}(k)\rightarrow\frac{\hbar
k^{2}}{2m\omega_{p}}\coth(\frac{\beta
\hbar\omega_{p}}{2}).\label{a.15}%
\end{equation}
This is the exact perfect screening behavior \cite{Martin99} which
is preserved by the RPA. Since $S^{(0)}(0)$ is finite at finite $t$
and vanishes as $k$ for $t=0$,
\begin{align}
\Delta\left(  r\right)    &
=\frac{1}{n}\int\frac{d\mathbf{k}}{\left( 2\pi\right)
^{3}}e^{-i\mathbf{k\cdot r}}\left(  \frac{1}{S^{RPA}(k)}-\frac
{1}{S^{(0)}(k)}\right)  \nonumber\\
& \rightarrow\frac{m\omega_{p}}{2\pi n\hbar\coth(\frac{\beta\hbar\omega_{p}%
}{2})}\int\frac{d\mathbf{k}}{\left(  2\pi\right)  ^{3}}e^{-i\mathbf{k\cdot r}%
}\frac{4\pi}{k^{2}}\nonumber\\
&
=\Gamma_{e}(t,r_{s})r^{\ast-1},\hspace{0.2in}\Gamma_{e}(t,r_{s})\equiv
\frac{2}{\beta\hbar\omega_{p}\coth(\frac{\beta\hbar\omega_{p}}{2})}%
\Gamma,\label{a.16}%
\end{align}
where $\Gamma=\beta q^{2}/r_{0}$ is the classical Coulomb coupling
constant. This is the result of eq. (\ref{5.12}).

\end{document}